\documentclass{article}

\usepackage{graphicx}
\usepackage{amsmath,amssymb}
\usepackage{csquotes}

\usepackage{cite}

\begin{document}

\title{Enforcing energy balance in coherently superimposed optical vortices}

\author{Jaime~Cisternas, Jaime~A.~Anguita, Gustavo~Funes \\
Facultad de Ingenier\'{\i}a y Ciencias Aplicadas, Universidad de los Andes,\\
Mons.\ Alvaro del Portillo 12455, Las Condes, 7620001 Santiago, Chile \\
Millennium Institute for Research in Optics (MIRO), Chile}

\date{\today}

\maketitle

\begin{abstract}
The generation of optical beams with multiple, mutually-coherent orbital-angular-momentum (OAM) modes using phase gratings is analyzed from the perspective of energy distribution and radial mode composition. We show that phase gratings designed with equally-weighted Laguerre-Gauss (LG) modes will generate beams with uneven energy distribution among OAM components. This unwanted outcome cannot be corrected by adjusting the width of the illuminating beam. We propose a way to design phase gratings that will produce a uniform energy distribution among the constituent OAM states after illumination, while minimizing the content of high radial modes.
This method is based on a generalized definition for the LG modes that takes advantage of the freedom to select their radial scales.
\end{abstract}

\section{Introduction}
Spatial modes with orbital angular momentum (OAM) form an important class of transverse modes and have drawn significant attention for their capability of carrying information on their theoretically unbound state space and the availability of several methods of generation and detection of optical vortices \cite{Allen1992,Gibson2004,YaoPadgett2011,Wang2012,Wang2015}. 

Data aggregation could be achieved in unguided optical communications ---from optical interconnects to free-space communications--- by using OAM-carrying optical modes in a multiplexing arrangement \cite{Gibson2004, Anguita2008, Wang2012, Sun2016}, in which each state carries the information of a single channel. Data aggregation is also possible through signal modulation in the OAM-state space, by superimposing two or more states, drawn from a defined set, to create a multi-dimensional information symbol \cite{Rop2012, Anguita2014, Djordjevic2017}. Multi-dimensional OAM modulation could be made more robust than OAM multiplexing in the presence of channel distortions, by choosing OAM state combinations (and thus, information symbols) to increase the minimum distance of the set
if the characteristics of the communication channel are measured and taken into account. This is particularly meaningful in the context of quantum or classical communications over long, unguided channels in terrestrial and earth-to-satellite links.

Several techniques have been proposed to create optical vortices, including mode transformation, astigmatic mode conversion, spiral phase-plates, amplitude and phase gratings, and computer-generated holograms \cite{Heckenberg1992, Beijersbergen1993, Sueda2004, Guo2005, Gonzalez2006, Ando2009, Berkhout2010, YaoPadgett2011, Ruffato2017}. The most common generation method used in current experimental demonstrations consists on passing a zero-order Gaussian beam through a forked grating etched on a transparent material or programmed onto a reflective spatial-light modulator (SLM)  \cite{Wang2012}.


If the diffractive element contains a single phase dislocation with state $\ell$, the emerging beam (in the first diffraction order) is not a pure state ---due to the amplitude ambiguity of the diffractive element---, and may be expressed as a composition of Laguerre-Gauss (LG) modes with orbital state $\ell$ and different radial states $p$. The emerging beam will be  sensitive to the initial conditions of the illuminating beam (i.e., to its diameter and curvature). The electric field of LG modes have a complex rotating phase $\exp(i \ell \phi)$ and an amplitude that depends on the generalized Laguerre polynomial $L_p^\ell (\cdot)$. The latter creates an additional dependence on the radial index $p$ \cite{Allen1992} which cannot be fully specified in phase-only gratings, thus creating intensity profiles that appear as a collection of concentric rings \cite{Dennis2009}.
The physical meaning of this `forgotten' radial number $p$ has been the subject of recent research \cite{Karimi2014,Plick2015}.
Appropriate procedures for measuring the LG spectrum of a light beam have been proposed \cite{Qassim2014,Geneaux2017}.
But equally forgotten were the radial scales of the LG modes and the possibility of freely adjusting the scale of each orthogonal subspace of OAM states.

Energy balance among OAM states is critical to achieve good signal-to-noise ratio on each dimension forming the basis of an OAM-based signal modulation scheme in an optical communication link.
In this work, we show that in coherently superimposed OAM states using a diffractive element, the emerging energy is not evenly distributed over the constituent states if the element imposing the superposition is designed by adding the electric fields of LG components with uniform amplitude weights and equal design waists. By analyzing the components of the emerging electric field using numerical propagation simulations, we propose and evaluate two better grating designs:
one based on a recipe, and another based on a simple minimization method.
We foresee applications of multi-vortex beams with uniform energy distribution among component OAM states in optical communications, quantum cryptography, and optical manipulation.

\begin{figure}[tbh]
\begin{center}
\includegraphics[width=0.65\textwidth]{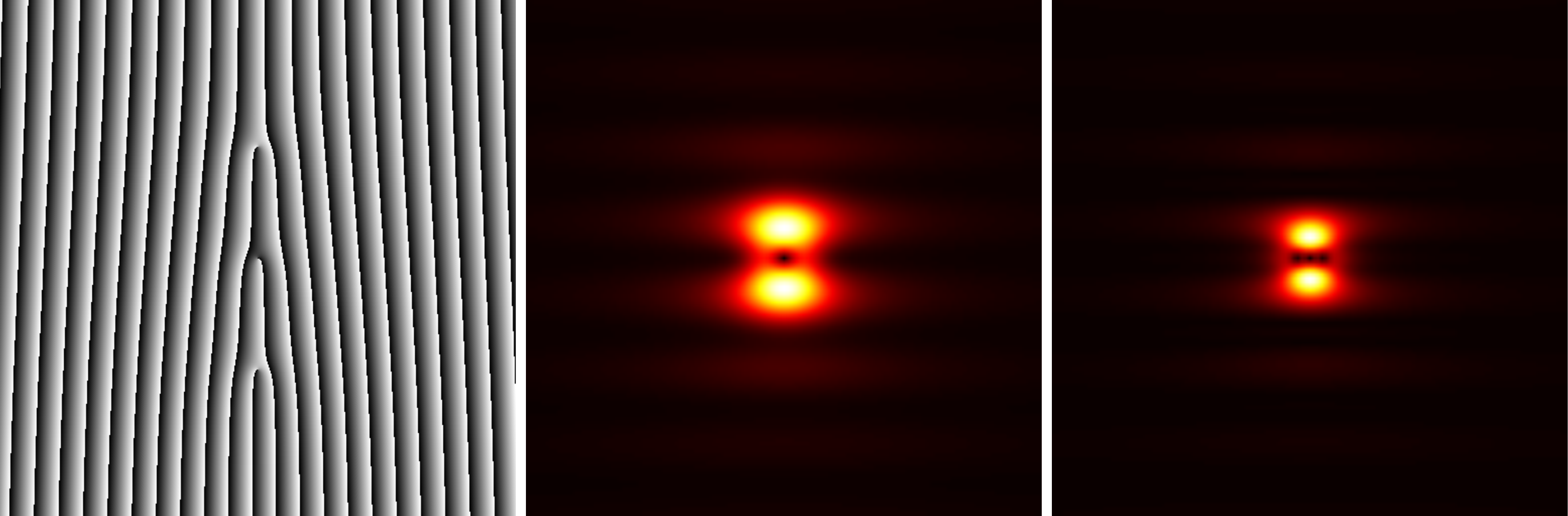}

(a) $\ell=\{1,3\}$ \\
\medskip

\includegraphics[width=0.65\textwidth]{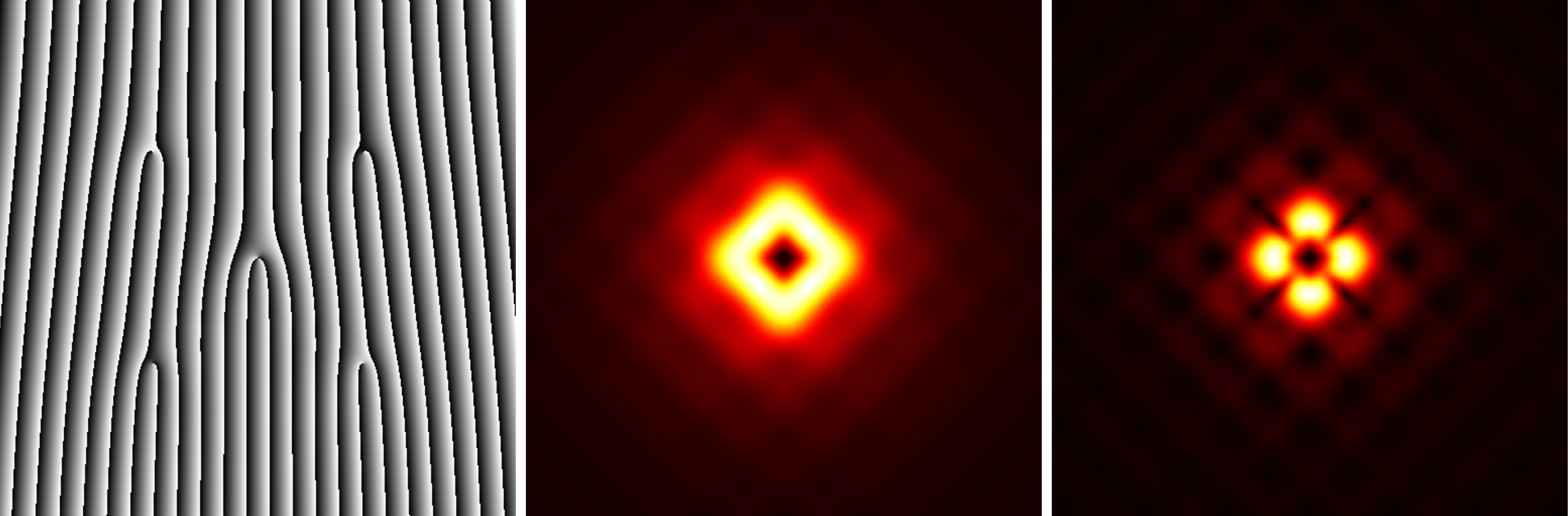}

(b) $\ell=\{2,6\}$
\end{center}
\caption{Examples of two-state OAM superpositions, using (a) $\ell=\{1,3\}$, and (b) $\ell=\{2,6\}$.
Each case shows a grating generated with equal design waists $w_0^\mathrm{ref}=1$ mm (left-most picture),
a far-field intensity with incident beam waist $w_{i}=1$ mm (central picture),
and a second far-field intensity with incident beam waist $w_{i}=2$ mm (right-most picture).
The sides of the square sections of the diffraction-grating are $5.12$ mm long.}
\label{fig:intpatt1}
\end{figure}

\section{Generalized LG basis for orbital state superpositions}

We refer to a coherent OAM state superposition as the coherent addition of two or more optical vortices' electric fields, each with distinct, integer OAM state, sharing the same optical axis. Unlike incoherent OAM superpositions, whose intensity patterns are concentric rings, the intensity profile of a coherent OAM superposition is not ring-shaped, and may take a great variety of distributions..

A conventional recipe to design a grating capable of diffracting a two-state OAM superposition is the following: extract the phase term of the electric field resulting from the addition of the complex amplitudes of two LG modes that use the same zero-order Gaussian waist diameter \cite{Anguita2014}. This bidimensional phase pattern can be used as a digital hologram if programmed to a SLM. Examples of two-state phase patterns are the forked gratings in Fig. \ref{fig:intpatt1}, where the gray scale represents phase values from $0$ to $2\pi$ radians.

A forked grating will show one or more off-center singularities, depending on the orbital states of the composition. For two states, the number of off-center singularities is equal to the absolute value of the orbital numbers' difference. This rule assumes that the design modes have the same zero-order waist. A detailed description follows.

Other techniques utilize the grating depth for controlling the amplitude of the beam with phase-only SLMs at the expense of a large fraction of laser power being diffracted towards unwanted orders \cite{Clark2016}. Here we restrict our study to \emph{blazed} gratings that basically diffract a single order. Now most of the ideas of our approach, such as the combinations of LG modes of different radial scales, can be combined with schemes that modify the amplitude.

Let $\Phi_{\ell,p}(r,\phi,z;w_\ell)$ be the electric field ---in cylindrical coordinates $r$, $\phi$, and $z$--- of a LG mode of orbital number $\ell$ (an integer which we also refer to as OAM state or topological charge), radial order $p$ (a nonnegative integer), and beam waist $w_\ell$. At a propagation distance $z$ from the beam waist, the electric field is given by

\begin{align}\label{eq:LGfield}
    \Phi_{\ell,p}(r,\phi,z;w_\ell)& \triangleq \sqrt{\frac{2p!}{\pi(p+|\ell|)!}} \,\frac{1}{w(\ell;z)} \left(\frac{r\sqrt{2}}{w(\ell;z)}\right)^{|\ell|} 
     L_p^{|\ell|} \left[\frac{2r^2}{w^2(\ell;z)}\right] \exp\left[\frac{-r^2}{w^2(\ell;z)}\right] \nonumber \\
    &\quad \times \exp(-i\ell\phi) \exp\left[\frac{-ikr^2 z}{2(z^2+z_R^2)}\right]
    \exp\left[i(2p+|\ell|+1)\tan^{-1}\frac{z}{z_R}\right] \:,
\end{align}
where $w(\ell;z)=w_\ell \sqrt{1+(z/z_R)^2}$ is the beam waist at distance $z$; $L_p^\ell(\cdot)$ designates the generalized Laguerre polynomial; $z_R=\pi w_\ell^2/\lambda$ is the Rayleigh range; $\lambda$ is the optical wavelength; and $k=2\pi/\lambda$ is the propagation constant.
In the previous definition one could also introduce an arbitrary translation $z_\ell$ along the propagation axis.

\subsection{Orthogonality of generalized LG modes}

The \emph{traditional} LG basis is defined using $w_\ell=w_0$, that is, a fixed spatial scaling for all modes.
It is well known that traditional LG modes are mutually orthogonal for distinct, integer values of $\ell$ and $p$. That is,
\begin{equation}\label{eq:orthogon}
\langle \Phi_{\ell_1,p_1}(r,\phi,z; w_0) , \Phi_{\ell_2,p_2}(r,\phi,z; w_0) \rangle = \delta_{\ell_1,\ell_2} \delta_{p_1,p_2}
\end{equation}
for any fixed value of $z$, where $\langle\cdot,\cdot\rangle$ is the inner product defined as the integration in the transverse plane,
\begin{equation}\label{eq:innerprod}
\langle \zeta(r,\phi) , \varphi(r,\phi) \rangle \triangleq \int_0^{2\pi} \int_0^\infty \zeta (r,\phi) \overline{\varphi} (r,\phi)\, r\, dr\, d\phi.
\end{equation}

Now the \emph{new} LG basis is defined by freely choosing the waists $w_{\ell}$ for each $\ell$. Orthogonality between modes in the new basis is maintained. That is,
\begin{equation}
\label{eq:orthogon2}
\langle \Phi_{\ell_1,p_1}(r,\phi,z;w_{\ell_1}),\Phi_{\ell_2,p_2}(r,\phi,z;w_{\ell_2})\rangle = \delta_{\ell_1,\ell_2} \delta_{p_1,p_2} 
\end{equation}

This can be verified considering two cases:
(i) if the two modes have the same angular momenta $\ell_1=\ell_2$ (and, by construction, $w_{\ell_1}=w_{\ell_2}$), orthogonality holds if $p_1 \ne p_2$, due to the orthogonality between the generalized Laguerre polynomials $L_{p_1}^\ell$ and $L_{p_2}^\ell$;
(ii) if the angular momenta are different $\ell_1 \ne \ell_2$, the inner product of the fields is zero, because the
angular integration of $\exp[-i(\ell_1-\ell_2)\phi]$ vanishes, regardless of the choice of $w_{\ell_1}, w_{\ell_2}$, $p_1, p_2$.
In the Appendices we review in more detail this derivation.

In contrast, Vallone~\cite{Vallone2017} studied the loss of orthogonality between two LG modes
of equal state $\ell$ but \emph{different} waists. The following inner product was considered:
$\langle \Phi_{\ell,p_1}(r,\phi,z;w_1),\Phi_{\ell,p_2}(r,\phi,z;w_2)\rangle$,
and expressed in terms of a hypergeometric polynomial.

LG bases constructed with a set of distinct values of $w_0, w_1, w_{-1}, w_2, w_{-2},\ldots$ will span the same space of superpositions of aligned vortices (as explained in the Appendices), even though their field components ---for any given $\ell$--- had different waists. Both traditional and new LG bases satisfy the \emph{completeness} requirement: any superposition of aligned vortices can be written as a finite or converging sum of LG modes, regardless of the set of $w_\ell$. However, choosing a `good' set can make convergence faster (i.e., with fewer terms).
The influence of basis scale on mode spectrum was also studied in \cite{Schulze2012, Vallone2017}.

We take advantage of the degrees of freedom provided by the beam waists $w_\ell$ in two ways: (i) in decomposing the reference field $u^\mathrm{ref}$ used to generate the grating and (ii) in decomposing the coherent superposition $u^\mathrm{s}$ generated by the grating, as illuminated by a zero-order Gaussian beam.

\begin{figure}[tb]
\begin{center}
\includegraphics[width=0.65\textwidth]{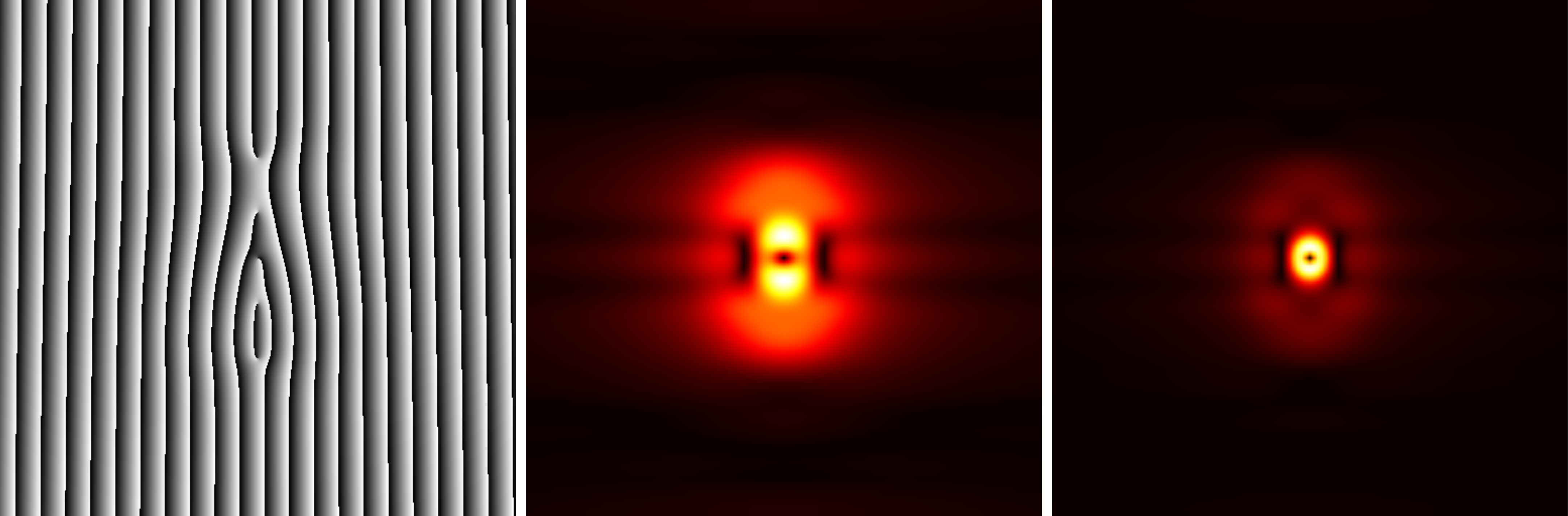}

(a) $\ell=\{1,3\}$ \\
\medskip

\includegraphics[width=0.65\textwidth]{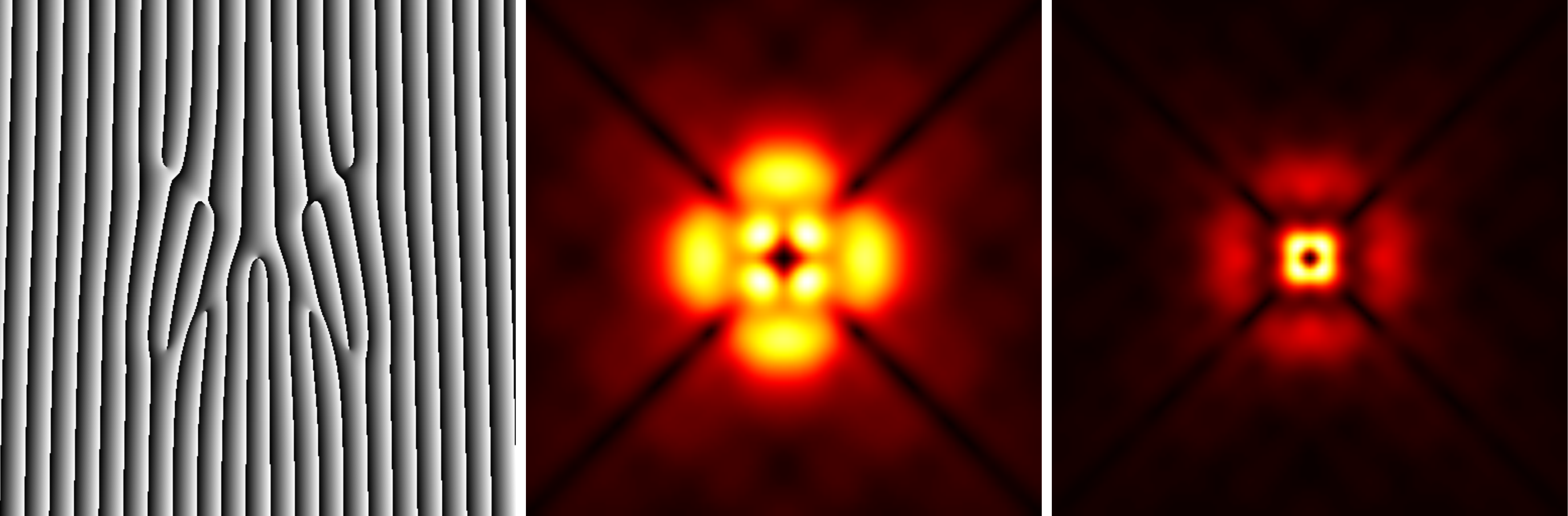}

(b) $\ell=\{2,6\}$
\end{center}
\caption{Examples of two-state OAM superpositions, using (a) $\ell=\{1,3\}$, and (b) $\ell=\{2,6\}$.
Each case shows a grating generated with equal design radii $r_\ell^\mathrm{ref}$ (left-most picture),
a far-field intensity with incident beam waist $w_{i}=1$ mm (central picture),
and a second far-field intensity with incident beam waist $w_{i}=2$ mm (right-most picture).
In (a) we used waists $w_1^\mathrm{ref}=1$ mm, $w_3^\mathrm{ref}=0.577$ mm,
and in (b) waists $w_2^\mathrm{ref}=1$ mm, $w_6^\mathrm{ref}=0.577$ mm.
The sides of the square sections of the diffraction-grating are $5.12$ mm long.}
\label{fig:intpatt2}
\end{figure}

\subsection{Reference field for the OAM superposition}

The reference complex field to be used as the interference pattern may be expressed as
\begin{equation}\label{eq:reference}
u^\mathrm{ref}(r,\phi) = \sum_{\ell \in \mathbb{Z}} c_{\ell}^\mathrm{ref} \Phi_{\ell,0}(r,\phi; w_\ell^\mathrm{ref}),
\end{equation}
in which we have chosen modes with $p=0$, $z=0$, and the set $\{c_{\ell}^\mathrm{ref}\}$ are arbitrary complex amplitude coefficients. The constituent LG fields $\Phi_{\ell,0}$ have distinct beam waists $w_{\ell}^\mathrm{ref}$. Choosing a set of distinct values for $\{w_\ell^\mathrm{ref}\}$ modifies the position of dislocations in the phase pattern and the energy distribution in the diffracted beam, even if $\{c_{\ell}^\mathrm{ref}\}$ were all equal. This is apparent by comparing the examples of superpositions $\{1,3\}$ and $\{2,6\}$ of Figs. \ref{fig:intpatt1} and \ref{fig:intpatt2}: equal design radius [$w_\ell^\mathrm{ref}=1$ mm in Fig. \ref{fig:intpatt1}] show a clearer separation of dislocations than the case with different design radius [$w_\ell^\mathrm{ref}=\{1,0.577\}$ mm in Fig. \ref{fig:intpatt2}]. The radius of the illuminating beam also impacts the diffracted mode: a simple observation of the gratings in both figures reveals that a certain minimum beam radius is required to ensure the illumination of all phase dislocations. 

If a certain LG superposition $u^\mathrm{s}$ is sought after at the far-field, a reference field $u^\mathrm{ref}$ and a proper incident beam radius must be chosen. As it will be shown later, the mode decomposition of $u^\mathrm{ref}$ and $u^\mathrm{s}$ are usually quite different, and the waists $\{w_\ell\}$ used for analysis may be quite different from the waists $\{ w_\ell^\mathrm{ref} \}$ of the reference field.

\subsection{Diffracted field of the OAM superposition}

The light field $u^s$ that emerges from the grating may be expressed as a superposition of mutually-orthogonal LG modes. That is,
\begin{equation}\label{eq:decomposition}
	u^\mathrm{s}(r,\phi,z) = \sum_{\ell \in \mathbb{Z}} \sum_{p \in \mathbb{Z}^\ast} c_{\ell,p}(w_\ell) \Phi_{\ell,p}(r,\phi,z; w_\ell),
\end{equation}
where the complex coefficient $c_{\ell,p}(w_\ell)$ ---for each $\ell$ and $p$ in the sum--- is given by the scalar projection
\begin{equation}\label{eq:c_coeff}
	c_{\ell,p}(w_{\ell}) = \langle u^\mathrm{s}(r,\phi,z), \Phi_{\ell,p}(r,\phi,z; w_\ell) \rangle,
\end{equation}
using definition given in Eq. (\ref{eq:innerprod}). The projection coefficients will not depend on $z$, as they are fully determined at the grating's exit surface. Thus, OAM components may be analyzed right after the grating on the corresponding diffraction order, even though the beam's intensity had not developed to its far-field distribution.

The fraction of energy observed on each LG component mode, represented by $|c_{\ell,p}(w_\ell)|^2$, will depend on the choice of the waists $w_\ell$. From the stand point of an analyzing device that observes $u^{s}$ without {\it a-priori} information, the choice of $w_\ell$ is arbitrary \cite{Schulze2012}.

An example of this important ---and somehow, counterintuitive--- characteristic, is described below.\\

{\it Example 1}

We numerically compare the energy distribution among OAM states $\ell=\{1,3\}$ in a beam $u^\mathrm{s}$ generated by two different digital forked gratings. The gratings are $2048\times 2048$ pixels wide, with spatial resolution $\Delta x=\Delta y=10$ $\mu$m,  average fringe pitch $\Lambda=250$ $\mu$m and wavelength $\lambda=660$ nm.
\begin{enumerate} 
\item Base case: a phase grating is designed with a reference $u^\mathrm{ref}_b$ using waists $w_1^\mathrm{ref}=w_3^\mathrm{ref}=1$ mm and equal amplitude weights, $c_1^\mathrm{ref}=c_3^\mathrm{ref}=1$.
\item A second grating with a reference $u^\mathrm{ref}_r$, in which $w_1^\mathrm{ref}=1$ mm. Waists $w_3^\mathrm{ref}$ is chosen such that the radius $r_\ell^\mathrm{ref} \triangleq w_\ell^\mathrm{ref}\sqrt{|\ell|/2}$ \cite{Padgett2015} satisfies the condition $r_1^\mathrm{ref}=r_3^\mathrm{ref}$. That is, the LG components would have intensity rings of equal radii, if they were generated independently.
\end{enumerate}

\begin{figure}[t]
\begin{center}
\includegraphics[width=0.60\textwidth]{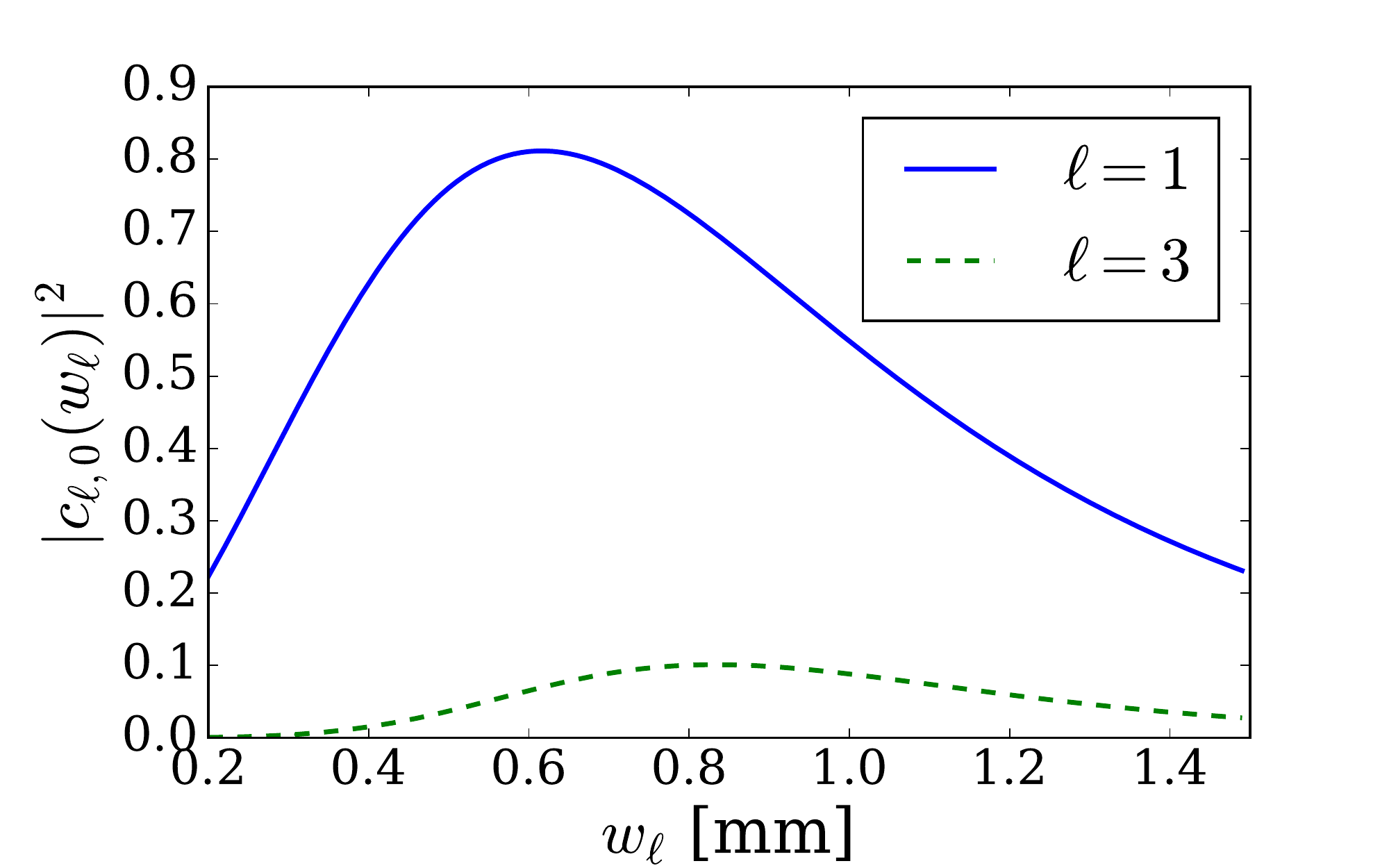}\\
(a)\\
\includegraphics[width=0.60\textwidth]{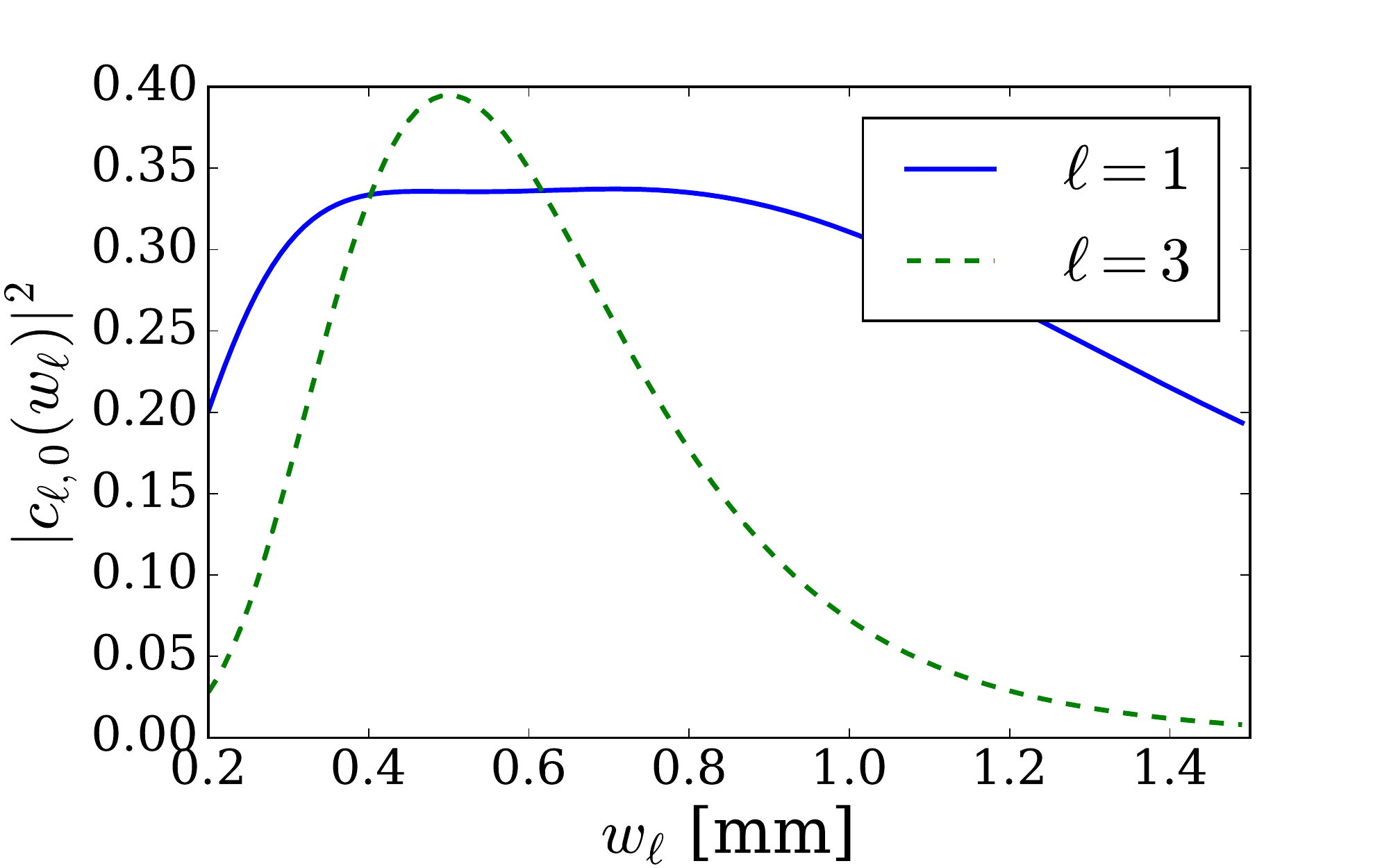}\\
(b)
\end{center}
\caption{Dependence of $|c_{\ell,0}|^2$ on the choice of waist $w_\ell$ in the observed superposition $u^\mathrm{s}$, for a grating programmed with states $\ell=\{1,3\}$. (a) $u^\mathrm{ref}$ uses equal design waists $w_1^\mathrm{ref}=w_3^\mathrm{ref}= 1$ mm. (b) $u^\mathrm{ref}$ uses waists $w_1^\mathrm{ref}=1\ \mathrm{mm}, w_3^\mathrm{ref}= 0.577\ \mathrm{mm}$. In both cases, the illuminating Gaussian beam has radius $w_\mathrm{i}=1$ mm.}
\label{fig:radial}
\end{figure}

Both gratings are illuminated using a beam width $w_\mathrm{i}=1$ mm and only the first diffraction order is analyzed. Fig. \ref{fig:radial} depicts the measured fraction of energy $|c_{\ell,0}(w_\ell)|^2$ (using Eq. (\ref{eq:c_coeff})) on the lowest radial order ($p=0$) as a function of $w_\ell$, for each $\ell$. Designs (i) and (ii) correspond to Fig. \ref{fig:radial} (a) and Fig. \ref{fig:radial} (b), respectively.

For both design cases, the measured energy varies significantly with $w_\ell$ and features a maximum. For the base case [Fig. \ref{fig:radial} (a)], a maximum of $|c_{1,0}|^2=0.811$ occurs at $w_1=0.62$ mm for $\ell=1$, and a maximum of $|c_{3,0}|^2=0.101$ at $w_3=0.83$ mm for $\ell=3$. This is obviously an unwanted outcome, as one component in the superposition has roughly 8 times the energy of the other. For the design with equal $r^\mathrm{ref}$ [Fig. \ref{fig:radial} (b)], maxima occur at  $w_1=0.710$ and at $w_3=0.50$ mm, respectively, and with a significantly better energy distribution. The values for all maxima and the values of $w_\ell$ at which they occur are listed in Table \ref{table:firstexample}.

\begin{table}[h]\centering
\caption{\emph{Example 1}. Fraction of energy on each diffracted OAM component (using Eq.(\ref{eq:c_coeff})) at their maximum values and their corresponding $w_\ell$ (given in mm).}
\label{table:firstexample}
\begin{tabular}{@{}lccccc}
\hline
 & \multicolumn{2}{c}{$\ell=1$} &  & \multicolumn{2}{c}{$\ell=3$} \\
 \cline{2-3} \cline{5-6}
 & $w_1$ & $|c_{1,0}|^2$ & ~~ & $w_3$ & $|c_{3,0}|^2$ \\
\hline
base case & 0.620 & 0.811 & & 0.83 & 0.101 \\
equal $r^\mathrm{ref}$ & 0.710 & 0.337 & & 0.50 & 0.395 \\
\hline
\end{tabular}
\end{table}

The generated light beam acquires the waist not from the grating but from the spatial distribution of the illuminating mode,
and the width follows $w_\ell = r_\mathrm{rms} \sqrt{2/(|\ell|+1)}$ \cite{Padgett2015}.
For a superposition of modes, the widths will depend on the location of phase singularities in the grating.

Using a different width on the illumination beam will produce a different set of plots, but the behavior described above prevails.\\

{\it Example 2}

A coherent superposition of three OAM states, namely, $\ell=\{2,6,10\}$, is made using the same numerical conditions of the first example. The gratings are defined as follows:
\begin{enumerate} 
\item Base case. Reference $u^\mathrm{ref}_b$ uses $w_2^\mathrm{ref}=w_6^\mathrm{ref}=w_{10}^\mathrm{ref}=1$ mm and equal amplitude weights. The grating is illuminated with a Guassian beam with $w_\mathrm{i}=1$ mm.
\item Reference $u^\mathrm{ref}_r$ uses $w_2^\mathrm{ref}=1$ mm and satisfies the condition $r_2^\mathrm{ref}=r_6^\mathrm{ref}=r_{10}^\mathrm{ref}$, as defined in \emph{Example 1}. Beam width $w_\mathrm{i}=1$ mm is used for illumination.
\end{enumerate}
Energy distribution and waists are detailed in Table \ref{table:secondexample}. Again, the equal-radii recipe gives a large improvement in energy distribution over the base case, although a significant contrast between the components still remains.

\begin{table}[h]\centering
\caption{\emph{Example 2}. Distribution of energy in a three-state superposition and the corresponding optimal waists (in mm).}
\label{table:secondexample}
\begin{tabular}{@{}lcccccccc}
\hline
 & \multicolumn{2}{c}{$\ell=2$} & & \multicolumn{2}{c}{$\ell=6$} & & \multicolumn{2}{c}{$\ell=10$} \\
  \cline{2-3} \cline{5-6} \cline{8-9}
 & $w^\mathrm{s}_2$ & $|c_{2,0}|^2$ & ~ & $w^\mathrm{s}_6$ & $|c_{6,0}|^2$ & ~ & $w^\mathrm{s}_{10}$ & $|c_{10,0}|^2$ \\
\hline
base case & 0.57 & 0.835 & & 0.76 & 0.0126 & & 0.85 & 0.0005\\
equal $r^\mathrm{ref}$ & 0.41 & 0.573 & & 0.46 & 0.0603 & & 0.41 & 0.149\\
\hline
\end{tabular}
\end{table}

Examples 1 and 2 show that to evaluate the merits of different phase gratings, it is key to identify
the scales $w_\ell$ of the generated OAM states and define the analysis LG basis accordingly
so most of the energy is concentrated on the lowest-order radial mode for every $\ell$.
This choice is also used to seek optimal design waists to equalize the energy of the coherent OAM superposition, as it is described in the following section.

\section{Grating design for equal-energy superpositions}

Acknowledging the impact of the observation basis on energy distribution, an independent metric needs to be used to analyze the generated modes. We define the \emph{eigen waist} of mode $\ell$ present in the diffracted beam $u^\mathrm{s}$ as
\begin{equation}\label{eq:properw}
w_\ell^\mathrm{s} \triangleq \mathrm{arg} \max_{w_\ell \in \mathbb{R}^+} |c_{\ell,0}(w_\ell)|^2,
\end{equation}
that is, we select the value of $w_\ell$ that delivers the maximum contribution of $|c_{\ell,0}(w_\ell)|^2$ for each orbital mode in $u^\mathrm{s}$, as computed with Eq.~(\ref{eq:c_coeff}). This choice is justified by the indetermination of the basis that analyzes $u^\mathrm{s}$ in terms of $p$.
A similar approach was used in Ref.~\cite{Vallone2017} but with a different goal in mind: to optimize the expansion of a generic beam in LG modes.

By adjusting the waists $\{w_\ell^\mathrm{ref}\}$ and the amplitudes $\{c_\ell^\mathrm{ref}\}$ of the superimposed orbital components of the grating ---as described by Eq. (\ref{eq:reference})--- one can design an `optimal' grating, in the sense that the diffracted beam satisfies the following criteria. Our goal is to balance the energy distribution among OAM states in the diffracted beam, while preserving similar eigen waists $w^s_\ell$, so that all constituent orbital modes be diffracted at a similar rate. The latter prevents increasing the optics size at the analyzer.

We propose the following optimization problem for a two-state superposition:
\begin{equation}
\begin{aligned}\label{eq:firstmin}
& \underset{w_{\ell_1}^\mathrm{ref}, w_{\ell_2}^\mathrm{ref}, c_{\ell_1}^\mathrm{ref}, c_{\ell_2}^\mathrm{ref}} {\text{minimize}} \,\,\,  (|c_{\ell_1,0}| -|c_{\ell_2,0}|)^2 + \eta (w_{\ell_1}^s- w_{\ell_2}^s)^2 \\
& \text{subject to:}\\
& (i) \,u^\mathrm{ref}=c_{\ell_1}^\mathrm{ref} \Phi_{\ell_1,0}(\cdot; w_{\ell_1}^\mathrm{ref}) + c_{\ell_2}^\mathrm{ref} \Phi_{\ell_2,0}(\cdot; w_{\ell_2}^\mathrm{ref}) \\
& (ii)\text{ $u^\mathrm{s}$ as generated by grating built from $u^\mathrm{ref}$} \\
& (iii)\text{ $w^\mathrm{s}_{\ell_1},w^\mathrm{s}_{\ell_2},c_{\ell_1,0},c_{\ell_2,0}$ extracted from $u^\mathrm{s}$.}
\end{aligned}
\end{equation}
The first term of the objective function seeks to balance energy among states and the second seeks to maintain similar diffraction angles (to constrain the optics). The constant $\eta>0$ is used as a normalization factor, to account for the differences in magnitude of $w_\ell^s$ and $c_{\ell,0}$.
We set $\eta=10^5$ in the examples presented in the article.
A smaller value of $\eta$ would produce a closer similarity between the energies of the modes, but at the cost of a larger difference between their waists.

Similarly, for $N$ states one can pose the following problem:
\begin{equation}
\begin{aligned}\label{eq:secondmin}
& \underset{ \{ w_{\ell_k}^\mathrm{ref}, c_{\ell_k}^\mathrm{ref} \}_{k=1\ldots N}} {\text{minimize}}
\sigma( \{ |c_{\ell_k,0}| \}_{k=1\ldots N} ) + \eta \sigma( \{ w_{\ell_k}^s \}_{k=1\ldots N} ) \\
& \text{subject to: } (i)\ u^\mathrm{ref}=\sum\nolimits^{N}_{k=1} c_{\ell_k}^\mathrm{ref} \Phi_{\ell_k,0}(\cdot; w_{\ell_k}^\mathrm{ref}) \\
& (ii)\text{ $u^\mathrm{s}$ as generated by grating built from $u^\mathrm{ref}$,} \\
& (iii)\text{ $\{w^\mathrm{s}_{\ell_k},c_{\ell_k,0} \}$ extracted from $u^\mathrm{s}$,}\\ 
&\text{where $\sigma(\{x_k\})$ is the variance of numbers $x_k$}.
\end{aligned}
\end{equation}

\begin{table}[h]\centering
\caption{Grating's optimal design parameters $\{w^\mathrm{ref}_\ell\}$  (in mm) and $\{c_\ell^\mathrm{ref}\}$ to reach a balanced state energy in $u^\mathrm{s}$, assuming $w_i=1$ mm.}
\label{table:optimal_grating}
\begin{tabular}{@{}lcccccccc}
\hline
Superp. & $w^\mathrm{ref}_{\ell_1}$ & $c^\mathrm{ref}_{\ell_1}$ & $w^\mathrm{ref}_{\ell_2}$ & $c^\mathrm{ref}_{\ell_2}$ & $w^\mathrm{ref}_{\ell_3}$ & $c^\mathrm{ref}_{\ell_3}$ & $w^\mathrm{ref}_{\ell_4}$ & $c^\mathrm{ref}_{\ell_4}$ \\
\hline
\{1,3\} & 0.95 & 1.0 & 0.58 & 1.0\\
\{2,6\} & 0.79 & 1.0 & 0.41 & 0.86  \\
\{2,6,10\} & 0.62 & 1.0 & 0.32 & 0.97 & 0.36 & 1.15 \\
\{1,5,9\} & 0.11 & 1.0 & 0.41 & 0.98 & 0.37 & 1.07 \\
\{1,-2,4,-5\} & 1.0 & 1.0 & 0.72 & 1.08 & 0.50 & 0.93 & 0.47 & 0.98 \\
\hline
\end{tabular}
\end{table}

\begin{table}[h]\centering
\caption{Eigen waists and energies obtained from the optimal grating designs  of Table \ref{table:optimal_grating}, illuminated with $w_i=1$ mm.}
\label{table:optimal_energy}
\begin{tabular}{@{}lcccccccc}
\hline
Superp. & $w^\mathrm{s}_{\ell_1}$ & $|c_{\ell_1,0}|^2$ & $w^\mathrm{s}_{\ell_2}$ & $|c_{\ell_2,0}|^2$ & $w^\mathrm{s}_{\ell_3}$ & $|c_{\ell_3,0}|^2$  & $w^\mathrm{s}_{\ell_4}$ & $|c_{\ell_4,0}|^2$ \\
\hline
\{1,3\} & 0.51 & 0.61 & 0.51 & 0.61 \\
\{2,6\} & 0.31 & 0.59 & 0.37 & 0.57  \\
\{2,6,10\} & 0.26 & 0.48 & 0.27 & 0.48 & 0.38 & 0.48 \\
\{1,5,9\} & 0.87 & 0.47 & 0.32 & 0.45 & 0.37 & 0.46 \\
\{1,-2,4,-5\} & 0.25 & 0.41 & 0.49 & 0.39 & 0.41 & 0.41 & 0.48 & 0.39 \\
\hline
\end{tabular}
\end{table}

We solve  Eqs. (\ref{eq:firstmin}) and (\ref{eq:secondmin}) with the Nelder-Mead algorithm implemented in SciPy, a Python library.
Although the functional assumes a rather simple expression, the connection between the design parameters and
the beam features involves the numerical propagation of a beam into the far field.
The Nelder-Mead algorithm \cite{NelderMead} is a general-purpose multi-variable minimization scheme that does not require the specification of partial derivatives and is provided in standard numerical libraries.
Other similar algorithms should work as well.
In our numerical experiments we used a variety of selections for $\{ \ell_i\}_{i=1 \ldots N}$.
In these experiences 20 iterations were enough for reaching convergence with a relative tolerance of $10^{-5}$.
Digital gratings are generated using a fixed resolution of $\Delta x=10$ $\mu$m and grating pitch $\Lambda=250$ $\mu$m.
We fix the illumination beam waist to $w_i=1$ mm.

The optimal parameters found for several combination of states are listed in Table {\ref{table:optimal_grating}}. Table {\ref{table:optimal_energy}} presents the resulting eigen waists and energy fractions, as produced by the optimal gratings whose parameters are given in Table {\ref{table:optimal_grating}}.

In all cases evaluated, the minimization reached a solution that produced a diffracted beam $u^\mathrm{s}$ with a good compromise between energy balance and similarity of waists.
In contrast to the results of Tables \ref{table:firstexample} and \ref{table:secondexample}, the balance of energy at each eigen waist is almost perfect with the proposed optimization method. 

In some of the examples, the `optimal' energies of the modes are not exactly equal.
This is a result of using the functionals (\ref{eq:firstmin}) and (\ref{eq:secondmin})
that combine the similarity of the energies and the similarity of the waists.
The optimal superposition will show a compromise that can be fine-tuned by adjusting the parameter $\eta$.

\begin{figure*}[!ht]
\begin{center}
\begin{tabular}{cc}
\includegraphics[width=0.45\textwidth]{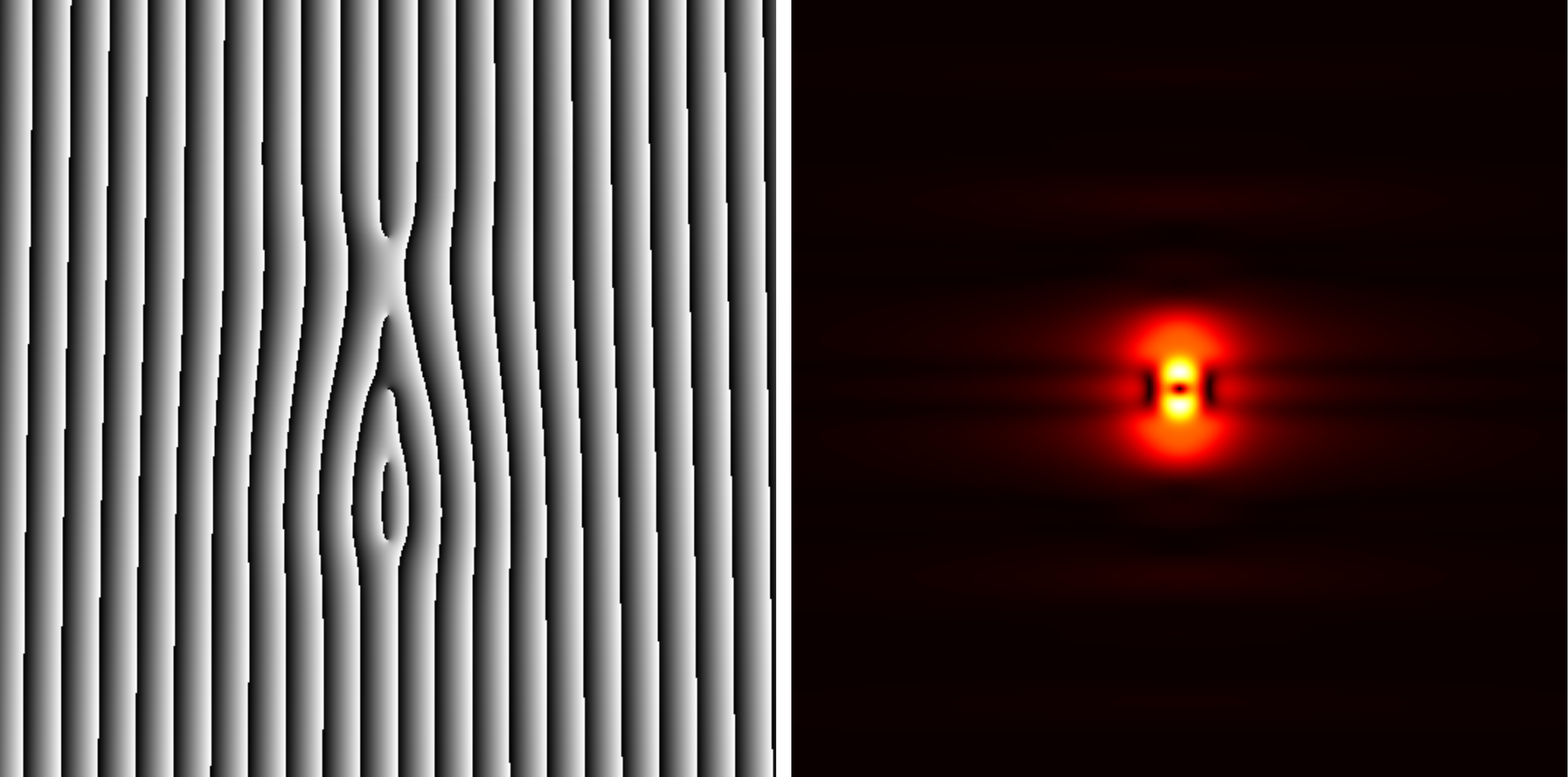} & \includegraphics[width=0.45\textwidth]{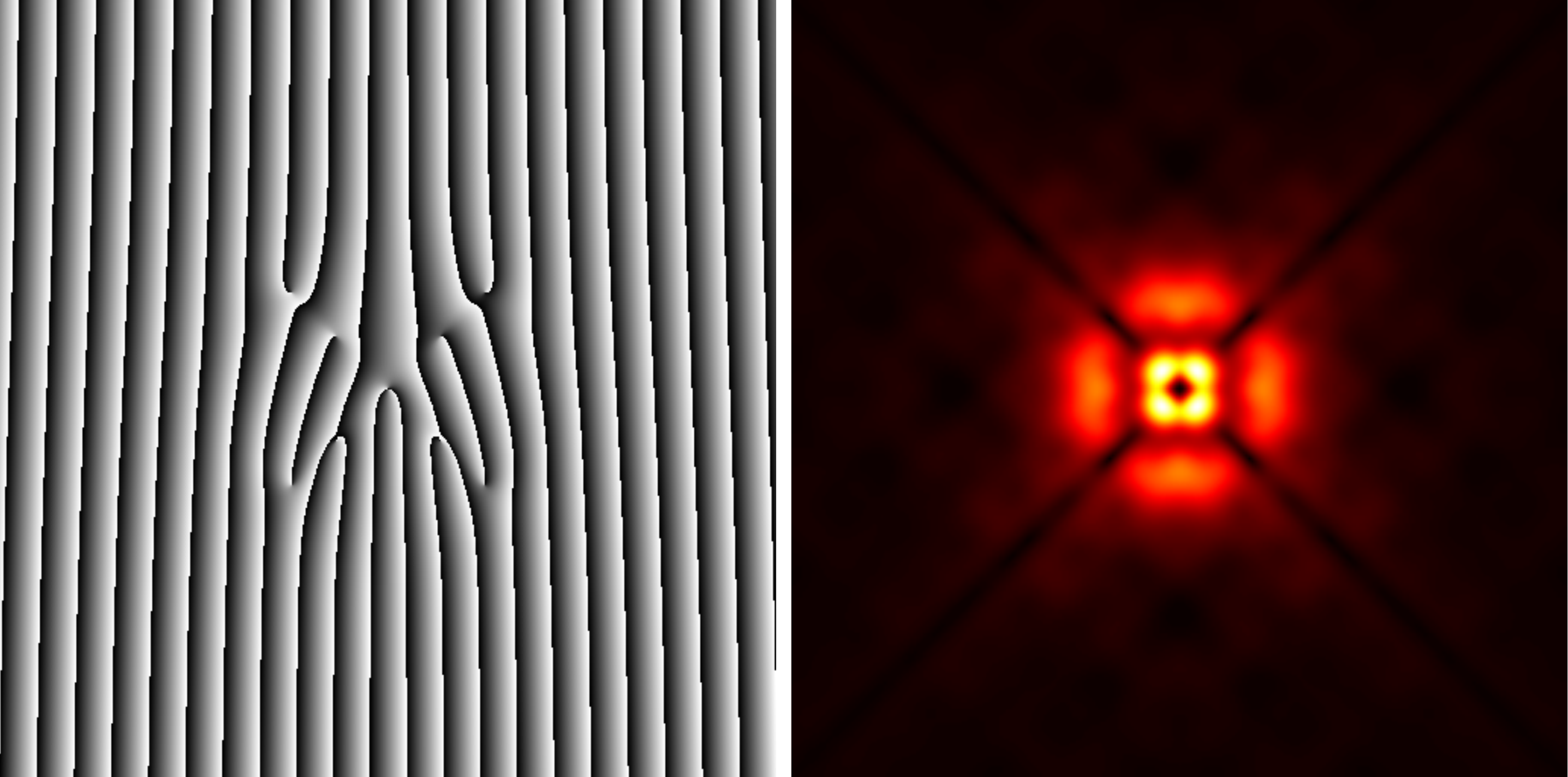} \\
(a) $\ell=\{1,3\}$ & (b) $\ell=\{2,6\}$ \\
\\
\includegraphics[width=0.45\textwidth]{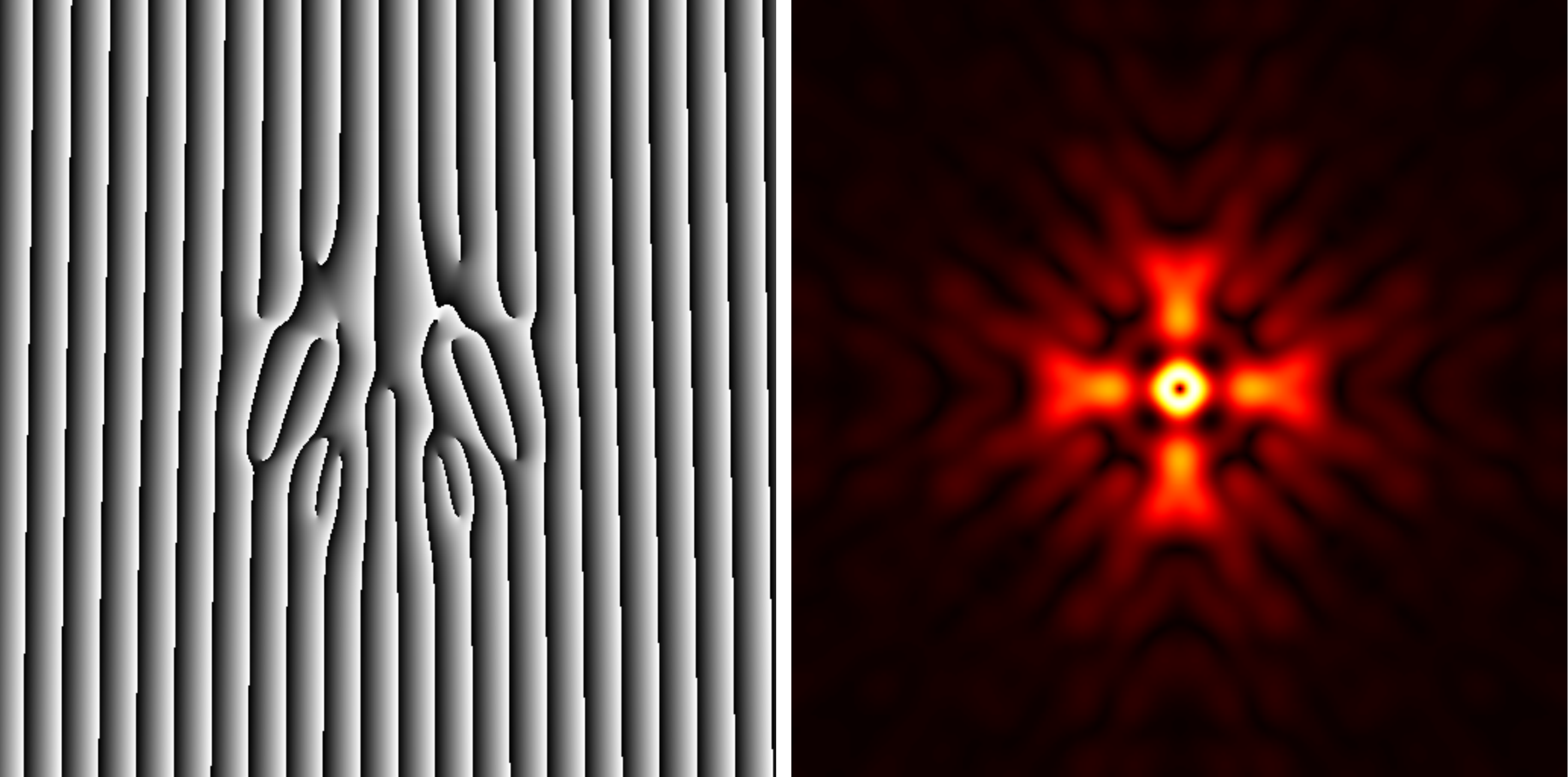} & \includegraphics[width=0.45\textwidth]{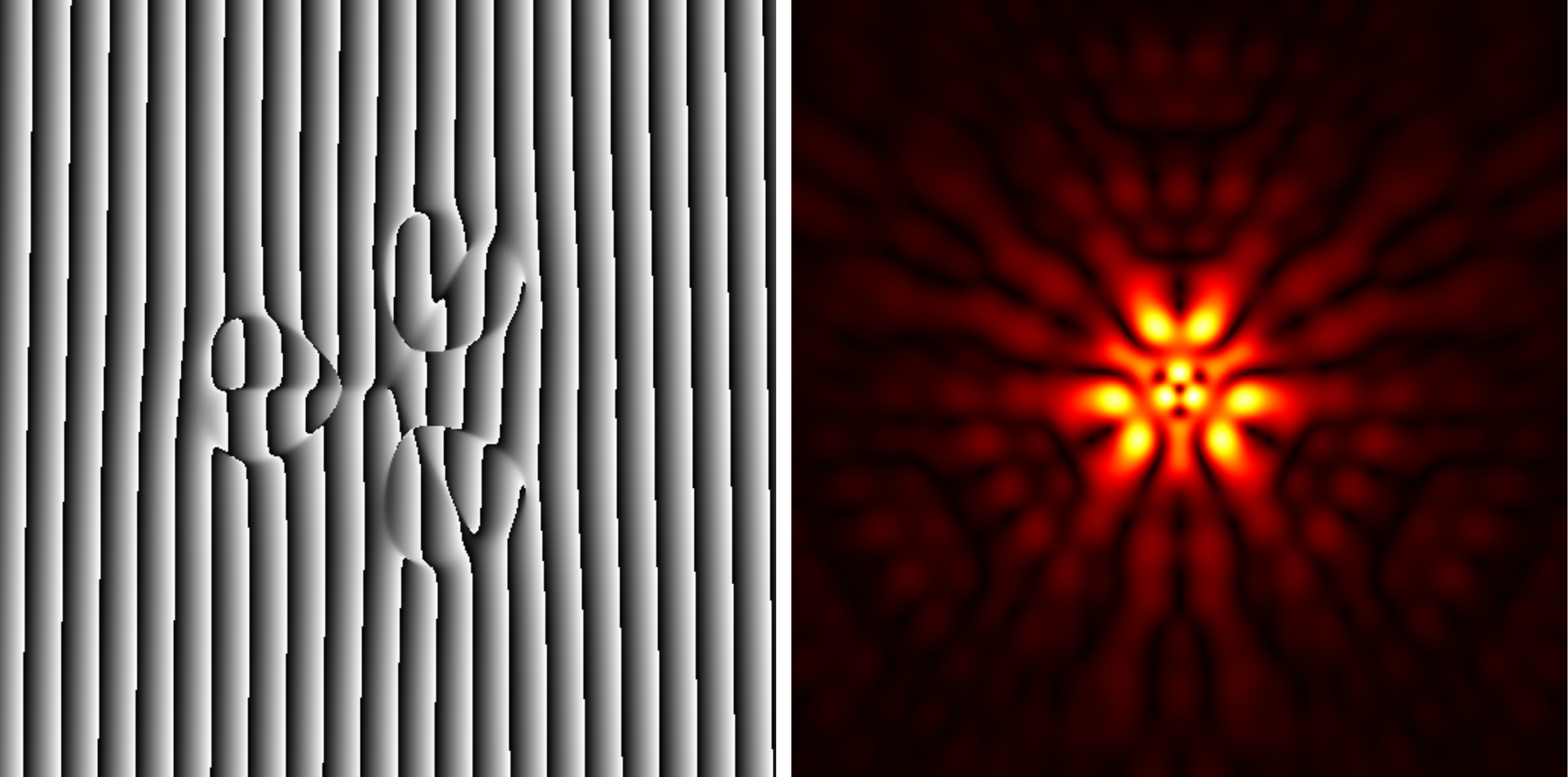} \\
(c) $\ell=\{1,5,9\}$ &  (d) $\ell=\{1,-2,4,-5\}$
\end{tabular}
\end{center}
\caption{Phase gratings and intensity profiles of optimal coherent superpositions.
The sides of the square sections of the diffraction-grating are $5.12$ mm long.}
\label{fig:best_gratings}
\end{figure*}

OAM states with negative sign may be added to the superpositions presented in Table {\ref{table:optimal_grating}} by simply using the waists and amplitudes of their positive counterparts. No further optimization is required. Figure \ref{fig:best_gratings} shows four sample intensity profiles as seen at the far field of $u^\mathrm{s}$ for superpositions $\{1,3\}$, $\{2,6\}$, $\{1,5,9\}$ and $\{1,-2,4,-5\}$ with balanced energy.

As a practical note, in any given system with predefined states, acquiring knowledge of the optimal $u^\mathrm{ref}$ for the grating and the eigen waists for the analyzer (e.g., the receiver in a communication system) is a one-time operation, and therefore, complexity is not an issue.

In this work our first motivation was to enhance the balance of energies of the OAM modes but keeping the waists also similar. This method is presented as a proof of principle: specific applications of OAM superpositions may lead to other requirements that can be mathematically represented in other definitions of the minimization functional (\ref{eq:secondmin}).

\section{Conclusions}

In this work we investigated the coherent superpositions of two or more coaxial LG modes
for use in multi-dimensional OAM-based modulation for optical communications. Particularly, we focused on the question of how to construct a coherent OAM superposition in which the constituent modes have equally distributed energy.
This energy balance is critical to achieve good signal-to-noise ratio on each dimension that forms the basis of an OAM-based modulation scheme in an optical communication link.
To answer this question, we generalized the definition of a LG basis, and then proposed three ways of manipulating the new degrees of freedom to create composite modes: (i) the basic scheme, that consists of a simple addition of the modes and creating a fork pattern with equal reference waists; (ii) the equal-radii scheme, that uses distinct reference waists in each constituent mode such that the radii of the rings are equal for each OAM mode; and (iii) an optimization scheme that selects the values of the reference waists to enforce energy balance on the diffracted beams.
We have shown that the first approach is flawed since it directs most of the energy into one state. The equal-radii recipe provides a moderate improvement in energy equalization. The optimization scheme provides an efficient algorithm that, for a given set of OAM states, constructs a phase grating that enforces a balanced energy distribution between the desired OAM modes.
Our future work involves an evaluation of the optimal gratings in a laboratory experiment.

\appendix

\section{Orthogonality and completeness of a generalized LG basis}

As we will see
the generalized LG modes defined in Eq.~(\ref{eq:LGfield}) are orthogonal to each other
under the inner product:
\begin{equation}\label{eq:innerprodapp}
\langle \zeta(r,\phi) , \varphi(r,\phi) \rangle \triangleq \int_0^{2\pi} \int_0^\infty \zeta (r,\phi) \overline{\varphi} (r,\phi)\, r\, dr\, d\phi.
\end{equation}
This definition satisfies the standard properties: it is linear in the first argument; antilinear in the second;
swapping of arguments is equivalent to complex conjugation; and it defines a norm:
\begin{equation}\label{eq:norm}
|| \zeta(r,\phi) ||^2 \triangleq \langle \zeta(r,\phi) , \zeta(r,\phi) \rangle
\end{equation}
that is always positive unless $\zeta(r,\phi)$ vanishes everywhere.

To prove the generalized orthogonality (Eq.~\ref{eq:orthogon2}), let us introduce the definition:
\begin{equation}\label{eq:simplif}
\varphi_{\ell,p}(r;w_\ell) \triangleq
\sqrt{\frac{4p!}{(p+|\ell|)!}} \,\frac{1}{w_\ell} \left(\frac{r\sqrt{2}}{w_\ell}\right)^{|\ell|} L_p^{|\ell|} \left[\frac{2r^2}{w^2_\ell}\right] \exp\left[\frac{-r^2}{w^2_\ell}\right]
\end{equation}
so every LG mode (we use $z=0$ here for simplicity) can be written as a product of a radial term and an angular term:
\begin{equation}
\Phi_{\ell,p}(r,\phi,0;w_\ell) =
\varphi_{\ell,p}(r; w_\ell)\ \frac{1}{\sqrt{2\pi}}\exp(-i\ell\phi)
\end{equation}
The properties of each factor are standard results of Sturm-Liouville theory that we summarize here.
Complex exponentials (for integers $\ell,\ell_1,\ell_2$) verify:
\begin{equation}\label{eq:orth1}
\int_{-\pi}^\pi e^{-i\ell_1 \phi} e^{i\ell_2 \phi}\ d\phi = 2 \pi \delta_{\ell_1 \ell_2}
\end{equation}
and:
\begin{equation}\label{eq:comp1}
\sum_{\ell \in \mathbb{Z}} e^{i\ell \phi} e^{-i\ell \phi'} = 2 \pi \delta(\phi-\phi')
\end{equation}
Generalized Laguerre polynomials (for nonnegative integers $\ell$ and $p_1,p_2$) verify:
\begin{equation}\label{eq:orth2}
\int_0^\infty e^{-x} x^\ell L_{p_1}^\ell (x) L_{p_2}^\ell(x) dx = \frac{(p_1+\ell)!}{p_1!} \delta_{p_1 p_2}
\end{equation}
and
\begin{equation}\label{eq:comp2}
\sum_{p \in \mathbb{Z}^\ast} e^{-x/2} x^{\ell/2} L_{p}^\ell (x) e^{-y/2} y^{\ell/2} L_{p}^\ell(y) = \frac{(p+\ell)!}{p!} \delta(x-y)
\end{equation}
These last two expressions can be rewritten using the definition of $\varphi_{\ell,p}(r;w_\ell)$ and the change of variable $x=2r^2/w_\ell^2$:
\begin{equation}\label{eq:orth3}
\int_0^\infty \varphi_{\ell,p_1}(r;w_\ell) \varphi_{\ell,p_2}(r;w_\ell)\, r\, dr = \delta_{p_1 p_2}
\end{equation}
and
\begin{equation}\label{eq:comp3}
\sum_{p \in \mathbb{Z}^\ast} \varphi_{\ell,p}(r;w_\ell)\ \varphi_{\ell,p}(r';w_\ell) = \frac{1}{r'} \delta(r-r')
\end{equation}
where the scale $w_\ell$ is completely arbitrary.
The generalized orthogonality Eq.~(\ref{eq:orthogon2}) now follows from Eqs.~(\ref{eq:orth1}) and (\ref{eq:orth3}).

The relevance of the completeness relations Eqs.~(\ref{eq:comp1}) and (\ref{eq:comp3}) can be explained by a simple argument.
For any function $f(r,\phi)$:
\begin{align}
f(r,\phi) &= \int_{-\pi}^\pi \int_0^\infty f(r',\phi') \delta(r-r')\delta(\phi-\phi')\, dr'\, d\phi' \nonumber \\
 &= \sum_{\ell,p} \langle  f(r',\phi'), \Phi_{\ell,p}(r',\phi';w_\ell) \rangle\ \Phi_{\ell,p}(r,\phi;w_\ell)
\end{align}
So if one can construct a delta function by adding products of LG modes,
then one can express any superposition of aligned vortices as a sum of LG modes.


\section{Equivalence between LG bases}

Considering two LG bases, one defined with a single scale $w$, and the other defined with $w_0, w_1$, $w_{-1}, w_2, w_{-2},\ldots$,
we can write for any function $f$:
\begin{equation}
f(r,\phi) = \sum_{\ell,p} a_{\ell,p} \Phi_{\ell,p} (r,\phi;w) = \sum_{\ell,p} b_{\ell,p} \Phi_{\ell,p} (r,\phi;w_\ell)
\end{equation}
and it is possible to connect the coefficients $a$ and $b$ by a linear transformation:
\begin{equation}
b_{\ell,p} = \sum_{p'} c^\ell_{p,p'} a_{\ell,p'}
\end{equation}
where the elements of the infinite transformation matrix:
\begin{equation}
c^\ell_{p,p'} = \langle \Phi_{\ell,p'} (r,\phi;w), \Phi_{\ell,p} (r,\phi;w_\ell)  \rangle
\end{equation}
There is no connection between coefficients with different $\ell$:
the change of basis is \emph{reducible}.
Ref.~\cite{Vallone2017} provides a closed expression for $c^\ell_{p,p'}$ in terms of the hypergeometric polynomial ${}_2 F_1$
with arguments that involve $\ell, p, p'$ and the ratio $w/w_\ell$.

The $L^2$-norm of the function $f$ can be computed as the sum of energies of the LG modes in either basis:
\begin{equation}
|| f||^2 = \langle f,f \rangle = \sum_{\ell,p} |a_{\ell,p}|^2 = \sum_{\ell,p} |b_{\ell,p}|^2
\end{equation}

Furthermore, if one introduces the projections (for any integer $\ell$):
\begin{equation}
f_\ell (r) \triangleq \int_{-\pi}^{\pi} \frac{1}{\sqrt{2\pi}} \exp(i \ell \phi) f(r,\phi)\, d\phi
\end{equation}
one can easily show:
\begin{equation}
f_\ell (r) = \sum_{p} a_{\ell,p} \varphi_{\ell,p}(r; w) = \sum_{p} b_{\ell,p} \varphi_{\ell,p}(r;w_\ell)
\end{equation}
and infer that the energy of each projection does not depend on the selection of the waists:
\begin{eqnarray}
||f_\ell ||^2 &=& \int_0^\infty f_\ell (r) \overline{f}_\ell (r)\, r\, dr \\
&=& \sum_{p} |a_{\ell,p}|^2 = \sum_{p} |b_{\ell,p}|^2
\end{eqnarray}
and
\begin{equation}
||f||^2 = \sum_\ell ||f_\ell ||^2
\end{equation}
These properties provide a mathematical framework for the freedom of scales of the LG modes.



\end{document}